# Generative Model-Based Ischemic Stroke Lesion Segmentation

Tao Song

East China Normal University
51151217005@stu.ecnu.edu.cn

**Abstract.** CT perfusion (CTP) has been used to triage ischemic stroke patients in the early stage, because of its speed, availability, and lack of contraindications. Perfusion parameters including cerebral blood volume (CBV), cerebral blood flow (CBF), mean transit time (MTT) and time of peak (Tmax) could also be computed from CTP data. However, CTP data or the perfusion parameters, are ambiguous to locate the infarct core or tissue at risk (penumbra), which is normally confirmed by the follow-up Diffusion Weighted Imaging (DWI) or perfusion diffusion mismatch. In this paper, we propose a novel generative model-based segmentation framework composed of an extractor, a generator and a segmentor for ischemic stroke lesion segmentation. First, an extractor is used to directly extract the representative feature images from the CTP feature images. Second, a generator is used to generate the clinical relevant DWI images using the output from the extractor and perfusion parameters. Finally, the segmentor is used to precisely segment the ischemic stroke lesion using the generated DWI from the generator. Meanwhile, a novel pixel-region loss function, generalized dice combined with weighted cross entropy, is used to handle data unbalance problem which is commonly encountered in medical image segmentation. All networks are trained end-to-end from scratch using the 2018 Ischemic Stroke Lesion Segmentation Challenge (ISLES) dataset and our method won the first place in the 2018 ischemic stroke lesions segmentation challenge in the test stage.

## 1    Introduction

Stroke is one of the primary causes of mortality and long-term disability worldwide [14]. Among all stroke, ischemic stroke accounts for 75-85%. Early diagnosis and treatment in the acute stage is very critical for the recovery of the stroke patient. CT perfusion (CTP) [1] is an important diagnostic method in ischemic stroke. It enables differentiation of salvageable ischemic brain tissue (penumbra) from irrevocably damaged infarcted brain (infarct core). Compared with CTP, magnetic resonance images (MRI) is more sensitive to the early parenchymal changes of infarction. But it can't be used in clinical scene extensively because of its drawback in availability, time and expenses. The quantitative perfusion parameters of CTP, are usually computed to identify the ischemic penumbra and the infarct core. The infarct core is defined as an area with prolonged MTT or Tmax, with markedly decreased CBF and CBV. The ischemic penumbra, which in most cases surrounds the infarct core, also has prolonged MTT or Tmax (typically > 6 seconds) [12]. But this pre-defined threshold does not take account of the variance in different patients.

In recent years, with the development of deep convolutional neural networks (CNN), segmentation in medical image processing has made great progress. CNN-based segmentation becomes much more accurate and effective than traditional feature or learning based methods. Criesan et al. [2] applied CNN to medical image segmentation task, which predicts a pixel's label based on information in a square window around itself separately. Later, Fully convolutional network (FCN)[3] is proposed, which can predict the image's pixel-label in a one-step forward operation. UNet[4] is a method based on FCN, which combines the localization and context information via an encoder-decoder structure and skip-connections. Specially, Nielsen et al. [10] introduced the CNN method in stroke segmentation task with a simple deep encoder-decoder structure.

On the other hand, generative model like GAN has thrived over the years. Xiang, et al. [11] proposed to generate from MR-T1 to CT image with a deep embedding CNN. Nie et al. [13] estimated CT image from MRI data with a 3D FCN. Loss functions like cross-entropy could introduce class-imbalance problems. Instead, dice loss function was proposed in [5] to alleviate this problem to some extent.

In this paper, we propose a generative model-based framework for ischemic stroke segmentation for CTP data, which consists of an extractor, a generator and a segmentor. The modified UNet-like structure, with encoder and decoder, is respectively used in extractor, generator and segmentor. This framework is trained and tested on ISLES 2018, and achieved state-of-the-art segmentation results in this challenge. The major contribution of this work is:

1) A generative-model based end-to-end segmentation framework, including an extractor, a generator and a segmentor, which is tailored for the multimodal imaging problem of ischemia stroke by generating a prediction of the follow-up DWI images.

2) An extractor to extract representative feature vectors directly from the dynamic CTP images, which is more suitable for learning from scratch than the traditional CT perfusion parameters.

3) A novel pixel-region loss function that combines both weighted cross entropy and generalized dice. It can stabilize the training process by balance the gradients of foreground and background samples.

The rest of the paper is organized as follows. Section 2 describes components of our generative model-based segmentation framework, followed by results and discussions in Section 3 and our conclusion in Section 4.

## 2    Method

### 2.1    Overall Framework

The overall pipeline of our method contains three parts, an extractor, a generator and a segmentor, as shown in Fig 1. The extractor is designed to produce the representative image or the most important information from CT perfusion images; The generator is to generate the follow-up DWI images using the output of the extractor and the perfusion parameters, which provides a more accurate modality for the segmentor; Finally, the segmentor is to precisely predict the ischemic stroke lesion using the generated data. All networks are trained end-to-end, and the infarct core will be automatically predicted in the inference phase.

Usually the traditional perfusion parameters, computed from original CT perfusion image, are ambiguous to identify the infarct core, while the infarct core can be recognized more accurately as hyperintense regions on the follow-up DWI images. So in this generative model-based segmentation framework, a generator is designed to generate DWI image from CT Perfusion data as well as traditional perfusion parameters. In order to generate realistic DWI images with more comprehensive input, we proposed an extractor network to extract representative feature images from the CTP images directly.

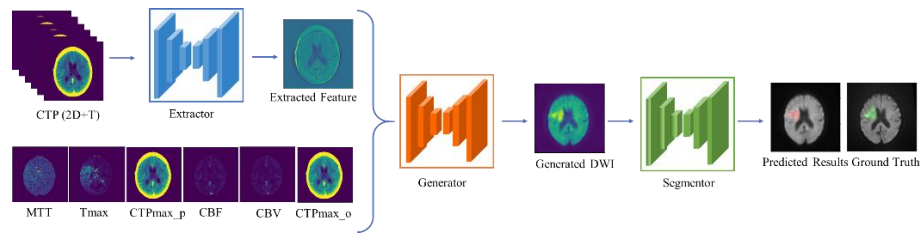

**Fig. 1. The overall pipeline of o**ur generative model-based segmentation framework. Here, **_CTPmax_o_** and **_CTPmax_p_** are the representative feature vectors extracted from CTP data, respectively.

## 2.2 Extractor

Normally, the perfusion parameters are calculated from the original dynamic CTP using the deconvolution method; Instead, we try to extract deep information from CTP using deep learning guidance, by the following four steps.

1) firstly, CTP data are summed along three spatial dimensions except time dimension, resulting in a summed time-density curve, which is considered as arterial impact function(AIF) approximately; 2) secondly, the summed coarse curve is smoothed using a 1D filter with kernel size of 5; 3) thirdly, we can get three important time points in the smoothed curve, onset time of contrast enhancement, time to peak and time of end enhancement. Then a predefined image frames (6 frames) are sampled uniformly between this time interval; The first three steps are pre-processing steps to obtain important CTP images with sufficient information. 4) finally, the extractor network is used to extract representative feature images directly from the sampled image frames.

The extractor network is a small UNet that halves number of feature channels of UNet in the down-sampling and up-sampling stage. The output of extractor is a single channel feature map with the same spatial size of input, without time dimension, which is calculated before sigmoid activation in the final layer. Through the sigmoid activation, this single channel feature map indicated the confidence probability of each pixel to be foreground. The L1 distance loss function is used to regress the confidence probability and is expressed as

$$L_e = \alpha * \|p - y\|_1 \tag{1}$$

where $p$ and $y$ represent the confidence probability of prediction and ground-truth respectively.

This ingenious design of the extractor provides more details for subsequent generator, which is one of the major contributions in our work.

### 2.3 Generator

UNet [4], which is a fully convolutional neural network and uses skip connections to combine low-level feature maps with higher-level ones, was used as generator. The generator generated the follow-up DWI images using the CT perfusion parameters and the output of the extractor as inputs.

In order to generate the DWI images, a novel loss function is designed for the generator. This loss function has two parts: one is L2 loss between the generated DWI and original DWI; the other is the feature space distance between the generated DWI and the original DWI. The overall loss function can be written as

$$L_g = \beta * W * \left\|DWI_g - DWI_o\right\|_2 + \gamma * \left(\left\|DWI_g^F - DWI_o^F\right\|_2\right) \tag{2}$$

where the $DWI_g$ and $DWI_o$ are generated DWI and original DWI, respectively. $DWI_g^F$ and $DWI_o^F$ are the feature maps, extracted from generated DWI and original DWI using deep neural networks. The networks transform the generated DWI and original DWI to high-level feature maps, similar as perceptual loss, in order to calculate their distance in feature space. $W$ is the heat map of ground truth of infarct core, calculated by signed distance function (SDF) [9], as shown in Fig. 3. The heat map focuses on the area of interest to enhance contrast between infarct area and normal brain tissue. We also compare the generator with a GAN design, and find that our loss function performs better than the discriminator's loss of the GAN, which will be discussed in Section 4.

It is the generator that converts the ambiguous boundary of the lesion into hyperintense regions of the DWI images to facilitate subsequent segmentation tasks. Comparison between the generated DWI and the original DWI in the validation set is shown in Fig. 2. From our experiments, it is much easier to identify the lesion from the generated DWI image than from the noisy CT perfusion parameters.

### 2.4 Segmentor

The segmentor network is designed with a novel attention UNet. Compared with the original UNet, it has an embedding network, called squeeze-and-excitation networks (SE Block) [6], and the batch normalization is replaced with switchable normalization (SN) [7]. SE Block adaptively recalibrates channel-wise feature responses by explicitly modelling interdependencies between channels using attention mechanism, which is good for extracting more important features. And SN is used to learn batch-wise, channel-wise as well as spatial-wise normalizer weights for normalization, it is robust to a wide range of batch sizes. These techniques are beneficial to improve the segmentation accuracy of the segmentor.

Ischemic stroke lesion segmentation has the same data unbalance problem as other medical image segmentation tasks. Thus we designed a combined loss of generalized dice [8] and weighted cross entropy (CE) as the segmentor's loss function. The proposed loss function uses both pixel-level and region-level punishment together, called pixel-region loss (PR loss).

$$L_s = \delta * \{W * CE - \log(generalized\ dice)\} \tag{3}$$

In this pixel-region loss function, $W * CE$ is weighted pixel-wise classification loss and $generalized\ dice$ is the class re-balancing dice loss. $Generalized\ dice$ can balance the loss of the foreground and background according to the number of pixels. But it can't punish when the prediction does not have overlap with the foreground ground truth, so the weighted cross entropy loss $W * CE$ is introduced. The combination of weighted cross entropy and $generalized\ dice$ can complement each other so that better segmentation was given by segmentor. In order to balance the gradient size of weighted cross entropy and generalized dice, a log operator is performed on the generalized dice.

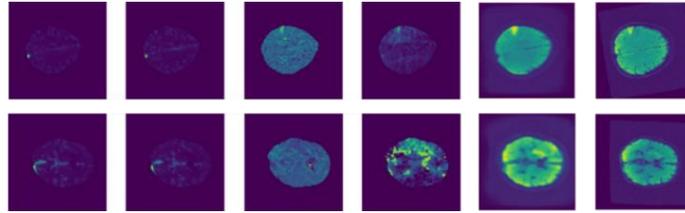

**Fig. 2.** Comparison of the Generated DWI with the original DWI, with 2 cases in the validation set shown. From left to right, the CBF, CBV, MTT, Tmax, generated DWI and original DWI.

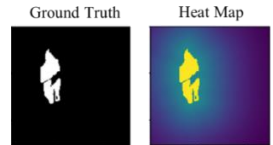

**Fig. 3.** The heat map of ground truth infarct core.

## 2.5 Training and Testing

As mentioned above, the extractor, generator and segmentor are the main components in our generative model-based segmentation framework. These networks are trained end-to-end from scratch using the ISLES 2018 training set. The training set is divided into four subsets to cross-validate the trained models using leave one out. Firstly, the extractor with input size of 256×256×6 pixels is used to extract the representative feature images. Then the extracted feature images and other perfusion maps are concatenated together to form an input image of with size of 256×256×7 pixels to the generator. Finally, the segmentor will predict the foreground region with the generated DWI of size of 256×256×1 pixels.

In the training stage, the weights of all networks are initialized using Xavier initialization and updated using RMSprop optimizer with batch size of 5. The warm-up strategy and step-by-step learning rate decay are used, and the learning rate is initialized at 0.002 and reduced by factor 0.2 after 180, 300 epochs, and the $\alpha$, $\beta$, $\gamma$ and $\delta$ we set at 1.0, 0.002, 1.2 and 1.0, respectively.

In the testing stage, we preprocess each case to extract CTP feature image, and concatenate the extracted feature images and perfusion parameters (CBF, CBV, MTT, Tmax) to feed the network of the generator, then the segmentor is used to predict the probability of infarct core using the generated DWI.

## 3 Experiments and Results

### 3.1 Dataset

The generative model-based segmentation framework is trained and tested using the 2018 Ischemic Stroke Lesion Segmentation Challenge dataset. 103 acute stroke patients from two centers were scanned with CT and CT perfusion within 8 hours of stroke onset time, and underwent MRI including DWI within 3 hours after CTP. Infarcted brain tissue can be recognized as hyperintense regions of the DWI images, and the ground-truth were manually drawn on DWI. Among the 103 patients, 60 are provided as training dataset, with CTP data, perfusion parameters (CBF, CBV, MTT, Tmax) as well as DWI data are given. The rest 43 are testing dataset, with only CTP data and perfusion parameters provided.

### 3.2 Results and Discussions

The generative model-based segmentation framework is implemented in PyTorch. In the training stage, all analysis of the trained model is performed on the cross-validation dataset. The detailed statistics are listed in Table 1, with a final state of art dice score of 62.40% achieved.

An ablation study to compare different techniques is shown in Table 1. Initially, a direct segmentation framework of our modified UNet using perfusion parameters (CBV, CBF, MTT, Tmax) as input and cross entropy as loss function, achieves a Dice score of 57.24%. Using the loss function proposed in section 2.4 with a combination of cross entropy and generalized dice, improves the Dice score to 58.33%, and makes the training process more stable. Revising the loss function for the segmentor to focus on the ROI heat map calculated by sign distance function, increases the Dice score to 59.30%. Meanwhile, an initial experiment using DWI in training (not available in testing phase) as input could achieve a Dice score of 87.56%, which inspired us to propose a two stage segmentation framework, which contains a generator and a segmentor.

In this framework, the extracted representative CTP image as well as perfusion parameters of CBF, CBV, MTT and Tmax are used as the input of the generator to generate a DWI image, then the generated DWI is used to predict the region of infarct core. It increase the dice score by 2 percentage on the original result with a segmentor only. It indicates that generator can convert the ambiguous boundary of the infarct core on perfusion images into clinically relevant DWI images to facilitate subsequent segmentation tasks. At the same time, we also tried to generate DWI with loss function of the discriminator to replace our generator's loss function, but the result is not as good as the generator's loss, is shown in Table I.

Next, we try to extract the deep information from CTP feature images to generate a more realistic DWI. Compared with previous result, using the extractor increase the result by one percentage. This shows that there is important deep information in the CTP feature images, which has not been fully exploited in the conventional CT perfusion parameters. This further inspires us in using deep learning to extract the traditional deconvolution-based perfusion parameter, which will be carried out in our future work.

As shown as Fig. 4, the predictions of our framework are compared with ground truths and predictions of a single U-Net based segmentor, also the generated DWI is compared with the original DWI. It demonstrates the effectiveness of our generative model-based segmentation framework.

**Table 1.** Ablation study of segmentation results using different methods. Here, SE and SN is the squeeze-and-excitation block and switchable normalization, respectively. And D is the discriminator's loss.

| Segmentor +SE+SN | Cross Entropy | General-ized Dice | Heatmap Weights | Generator | Extractor | Dice (%) |
|---|---|---|---|---|---|---|
| ✓ | ✓ | | | | | 57.24 |
| ✓ | ✓ | ✓ | | | | 58.33 |
| ✓ | ✓ | ✓ | ✓ | | | 59.30 |
| ✓ | ✓ | ✓ | ✓ | ✓(D) | | 60.44 |
| ✓ | ✓ | ✓ | ✓ | ✓ | | 61.49 |
| ✓ | ✓ | ✓ | ✓ | ✓ | ✓ | **62.40** |

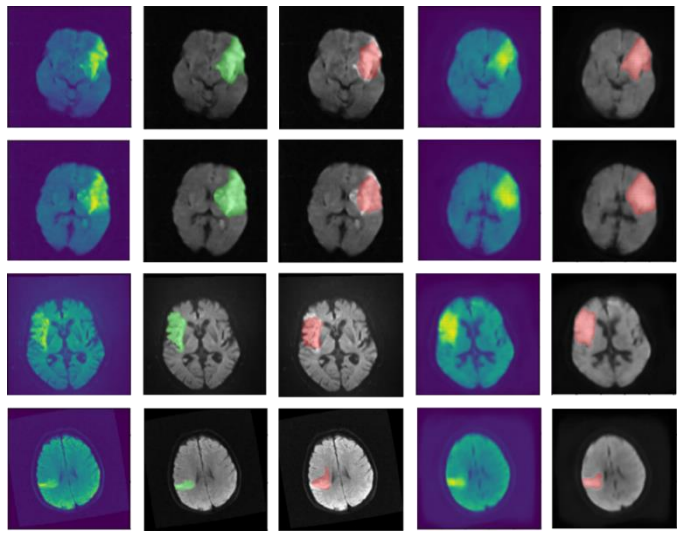

**Fig. 4.** Segmentation results of 4 cases in the validation set compared with the ground truth. The ground truths and predictions are given in green and red, respectively. From left to right, the original DWI, ground truth superimposed on original DWI, predictions of a single UNet segmentor superimposed on the original DWI, the generated DWI and the predictions superimposed on the generated DWI.

## 4       Conclusion

In this paper, a novel generative model-based end-to-end segmentation framework is proposed, which consists of an extractor, a generator and a segmentor, for ischemic stroke lesion segmentation. Results analysis shows that our framework is robust to the challenging problem of accurate stroke lesion segmentation using CT perfusion images. For this highly unbalanced segmentation problem, we also designed a novel pixel-region loss function. Our framework obtained a Dice coefficient of 62.40% in cross validation stage and won the first place in the test stage of 2018 ischemic stroke lesions segmentation challenge. In the future, we will explore the usage of this framework in clinical workflow of stroke patients, and possibly replace devolution based perfusion parameters with deep learning based perfusion feature images.